\documentstyle[preprint,aps,epsf,epsfig]{revtex}
\preprint{\vbox{
\hbox{UMN-TH-1538/97}
\hbox{astro-ph/9705107}}}      
\def\he#1{\hbox{${}^{#1}$He}}

\def\beq{\begin{equation}}
\def\eeq{\end{equation}}
\def\beginapjbib{\begingroup \section*{\large \bf References}
         \parskip=.5ex plus 1.0pt
         \def\bibitem{\par \noindent \hangindent\parindent
                \hangafter=1}}
\def\endapjbib{\par \endgroup}
\def\beginfig{\begingroup \section*{\large \bf Figure Captions}
         \parskip=.5ex plus 1.0pt
         \def\figitem{\par \noindent \hangindent\parindent
                \hangafter=1}}
\def\endfig{\par \endgroup}
\def\hii{H\thinspace{$\scriptstyle{\rm II}$}~}
\def\etal{{\it et al.}~}
\def\lsim{\lower0.6ex\vbox{\hbox{$ \buildrel{\textstyle 
<}\over{\sim}\ $}}}
\def\rsim{\lower0.6ex\vbox{\hbox{$ \buildrel{\textstyle 
>}\over{\sim}\ $}}}
\def\msun{${\,M_\odot}$}

\begin{document}
\title{A Bayesian Estimate of the Primordial Helium Abundance}
\author{Craig J. Hogan${}^1$,  
Keith A. Olive${}^2$ and Sean T. Scully${}^2$}

\address{${}^1$University of Washington \\
Astronomy and Physics Departments, Box 351580, Seattle, WA 98195-1580\\
${}^2$School of Physics and Astronomy,\\
University of Minnesota,
Minneapolis MN 55455 USA 
}
\maketitle
\vskip -.2in
\begin{abstract}
We introduce a new statistical method to estimate the 
primordial helium abundance $Y_p$ from observed abundances in
a sample of galaxies which 
have experienced   stellar helium enrichment.
Rather than using linear regression on metal abundance we
construct a likelihood function using a Bayesian prior, 
where the key assumption is that the true
helium abundance must always exceed the primordial value.
  Using a sample of   measurements compiled from
the literature we find   estimates of 
$Y_p$ between 0.221  and 0.236,
depending on the specific subsample and  
prior adopted,    consistent with previous
estimates   either from a linear extrapolation
of the helium abundance with respect to metallicity, or from the
helium abundance of the lowest metallicity \hii region, I Zw 18.
We also find
an upper limit which  is   insensitive
to the specific subsample or   prior, and
 estimate a model-independent bound
 $Y_p < 0.243$ at 95\% confidence, favoring 
a low cosmic baryon density and    a high primordial 
deuterium abundance. The main uncertainty is not
the model of stellar enrichment but possible common systematic
biases in the estimate of $Y$ in each individual HII region.
\end{abstract}
  

\section{Introduction}
Big Bang Nucleosynthesis (BBN) makes a  clean prediction  for the
primordial helium abundance   $Y_p$, depending on only one parameter
(the baryon-to-photon ratio $\eta$) and on that only weakly.  
A precise measurement of $Y_p$ is therefore  necessary
to test 
BBN  even in the present situation where
measurements of other predicted quantities constraining $\eta$, such as the
primordial lithium abundance, deuterium abundance and 
the  cosmic baryon density,
are not yet so precise  
(e.g. Walker \etal 1991,
Smith \etal 1993,  Sarkar 1996, Fields \etal 1996, Hogan 1997ab).

Even though the bulk of  the helium of the Universe originates in the Big Bang,
the additional helium enrichment by  stars cannot be ignored 
in estimating the primordial abundance 
from observations of present-day helium.
Unlike the case of deuterium, 
we do not have the option of measuring directly 
the  nearly primordial  abundance of \he4
at high redshift.
Lyman-$\alpha$ absorption by He$^+$ at
 high redshift now gives a rough direct estimate of
primordial abundance (Hogan, Anderson
\& Rugers 1997),
but require uncertain
 ionization corrections  to estimate $Y$.
 The high precision 
needed for strong
 cosmological tests can only be attained with high signal-to-noise
measurements of nebular emission lines in well characterized nearby
\hii regions (Pagel \etal 1992, Izotov, Thuan, \& Lipovetsky, 1994, 1997,
Skillman \etal 1994, 1997),
 necessarily requiring measurements in gas  with a certain amount
of stellar helium superimposed on the primordial helium.

Under these circumstances, what is the best way to
estimate the primordial helium abundance $Y_p$?
It certainly helps to find \hii regions with as little
stellar helium as possible, as indicated both by their
helium abundances and by the abundances of other heavier
elements. The best studied example is I Zw 18 (Pagel \etal 1992,
Skillman \& Kennicutt 1993, Izotov, Thuan, \& Lipovetsky, 1997), 
with an average estimated helium abundance from five independent 
measurements $Y=0.230\pm 0.004$ (Olive, Skillman, \& Steigman 1997),
and a metallicity 1/50th of solar. 
In the analysis below, we will use samples drawn from  the  
complete sample of 62 individual \hii regions
(including two in I Zw 18)    compiled
by Olive, Skillman, \& Steigman (1997) from the data in Pagel \etal (1992), 
Izotov, Thuan, \& Lipovetsky, (1994, 1996), and
Skillman \etal (1994, 1997).
This sample represents all published $Y$ measurements
  of \hii
regions with O/H $\le 1.5 \times 10^{-4}$ and N/H $\le 1.0 \times 10^{-5}$
(for comparison (O/H)$_\odot = 8.5 \times 10^{-4}$ and 
(N/H)$_\odot = 1.1 \times 10^{-4}$).

One statistical technique to extract a primordial helium abundance from
such a sample was introduced by Peimbert \& Torres-Peimbert (1974)
and is still widely used.
For each galaxy in the sample, a metallicity $Z$ (either
  O/H  or  N/H) and
helium abundance $Y$ are measured; one then fits 
a linear relation between $Y$ and $Z$ and extrapolates to 
zero metallicity to find the primordial value.
This technique   offers
insights into the nuclear evolution of these systems,
and for relatively large
enrichments the extrapolation is
 empirically well founded. In Olive, Steigman
\& Walker (1991) and in Olive \& Steigman (1995), it was shown that
these two quantities as measured in extragalactic
\hii regions are in fact strongly correlated. The stability of the resulting
fits was tested by a statistical bootstrap in Olive \& Scully (1996) 
and in Olive, Skillman, \& Steigman
(1997) showing that the results were not particularly sensitive to any 
individual data point. 

However, one of the limitations of this method is the need to
 assume
 a linear relation between $Y$ and $Z$, 
which is not well motivated.  For example, helium and oxygen
are not produced in the same stars; while helium is produced  primarily in 
intermediate mass stars, oxygen is produced only in stars with masses
$\rsim 8$\msun.
Physically realistic enrichment models often include quadratic terms,
as well as stochastic variations in enrichment history,
including different slopes $dY/dZ$.  It is not clear that complicating
the fits to include such effects results in a value for primordial   
$Y_p$ with any greater statistical significance or reliability.
Moreover most of the information on the primordial abundance is contained
in the lowest metallicity points, where the correlation is 
not very reliably established; this information is not 
being efficiently used in regression fits dominated
by highly enriched regions.

Another approach has been to simply take 
the lowest, best measured points and use them as estimates of
(or at least limits on) the primordial abundance. 
 Indeed, it has been argued that
it may be sufficient to determine the primordial \he4 abundance
from even a single well studied low metallicity \hii region such as I Zw 18
(Kunth \etal 1994), although to avoid bias it 
should  be  chosen on the basis of low
metallicity and not low helium abundance. Although
these approaches yield results
 consistent with the
linear fit to the data, 
caution is needed   to avoid subtle biases  if one starts off by selecting
a sample by  intentionally choosing the lowest points.
Moreover in a sample of more than just one point we need a 
statistical method to   
combine the data, which recognizes the real spread in 
stellar helium enrichment and still recovers an unbiased estimate
of the primordial value in spite of the obvious ``bias'' that 
 the nonprimordial contributions are always additive.

We explore here a simple
but systematic Bayesian approach to these issues,
which we believe is the simplest recipe to extract 
an unbiased primordial abundance estimate from the data,
free from detailed assumptions about metal enrichment.
 We  aim to assume as little as
possible---   only that 
there is some universal primordial helium abundance, and that
all subsequent evolution has increased the abundance.
We  encapsulate these assumptions
mathematically  with a Bayesian ``prior'' and
  derive  statistical
 constraints on the primordial abundance
which  explicitly recognize the bias
introduced by stellar enrichment.  Inspiration for our approach
and more background on the statistical methods is 
given  by Press' (1996) application of Bayesian arguments
to estimates of the Hubble constant.

\section{Method and Application}
Suppose we have a sample $S$ consisting of a series
of abundance measurements $Y_i$ in a set of galaxies.
In each  case $i$ there is some true abundance 
$Y_{iT}$ which we do not measure, because
of the measurement error.
We seek to evaluate the relative probability (or likelihood)
of obtaining the data in sample $S$, given
a primordial abundance $Y_p$:
\begin{equation}
{\cal L}(Y_p)\equiv P(S | Y_p)=\prod_i P(Y_i | Y_p)
\end{equation}
 where $P(Y_i | Y_p)$ denotes the probability of 
obtaining the measurement $Y_i$ given the 
primordial value $Y_p$. For this in turn we
write
\begin{equation}
 P(Y_i | Y_p)= \int dY_{iT}  P(Y_i | Y_{iT}) P(Y_{iT} | Y_p).
\end{equation}
The first term in the integrand is just the distribution of 
measurement errors in each case. 
For the second term we need to explicitly construct
a Bayesian prior $P(Y_{iT} | Y_p)$. This
``enrichment probability function''
 encodes our  assumptions
about what the distribution of helium abundances   in
galaxies in our sample ought to be, given a primordial abundance.

What do we know about $P(Y_{iT} | Y_p)$?
The most important property,
supported by all models of nuclear evolution,  is that
there is no net destruction of helium: stars 
can only increase the primordial abundance.
Thus nowhere can the true abundance
be less than $Y_p$: $P(Y_{iT} | Y_p)=0$ for $Y_{iT}\le Y_p$.
At a more detailed level, the shape of 
 this prior depends on what we think about the chemical
histories of the galaxies in the 
 sample under study; there will be some range
of $Y_p$ of width $w$ for which $P(Y_{iT} | Y_p)\neq 0$.
  Fortunately as we shall 
demonstrate, within reasonable limits the
detailed shape for $Y_{iT}\ge Y_p$
does not  much affect the statistical limits on $Y_p$: 
the most important thing is its asymmetry
about $Y_{iT}= Y_p$.

Within  this framework we can now explore  
estimates of $P(S | Y_p)$, and consequent statistical
constraints on $Y_p$. To start with, (1)  assume  that
 the observational errors  are normally distributed
with variances $\sigma_i^2$, and (2) assume  the   
simplest form for $P(Y_{iT} | Y_p)$, a uniform 
distribution or top-hat function of width $w$,
 $P(Y_{iT} | Y_p)= c (\Theta(Y_{iT} - Y_p) - 
\Theta(Y_{iT} - (Y_p + w)))$.
The  likelihood function is then given by
\begin{equation}
{\cal L}(Y_p) = \prod_{i=1}^{N} \int_0^1dY_{iT} 
{1 \over \sqrt{2 \pi} \sigma_i} e^{-(Y_i - Y_{iT})^2/2\sigma_i^2}
              P(Y_{iT} | Y_p),
\label{prob1}
\end{equation}
where the product is over the $N$ sample data points. 
The integral in (\ref{prob1})
can be performed analytically, resulting in
\begin{equation}
{\cal L}(Y_p) = \prod_{i=1}^{N} P(Y_i | Y_p)  =  \prod_{i=1}^{N}  
{c_i \over 2} \left\{ {\rm erf}\left( 
{Y_p + w - Y_i \over \sqrt{2} \sigma_i}\right)
 - {\rm erf}\left( {Y_p - Y_i \over \sqrt{2} \sigma_i}\right) \right\}
\label{prob2}
\end{equation}
The normalization constants $c_i$ are chosen so that each of the individual 
 functions $P(Y_i | Y_p) $, when integrated over $Y_p$, yield  unity.
To a very good approximation, $c = 1/w$.
For sufficiently large $w$, the first error function is approximately unity.

We choose to leave $w$ as a free parameter
and maximize the likelihood (\ref{prob2}) in both $Y_p$ and $w$---
that is, we estimate $w$ from the sample itself.
For each choice of $ (Y_p,w)$ we compute the likelihood ${\cal L}(Y_p,w)$;
the maximum of the likelihood gives the best values of these parameters.
The relative likelihood of other choices
then allow us to
derive constraints on $Y_p$ and $w$. 
This can be illustrated in 
contour plots  showing the   equal likelihood contours
representing 1, 2 and 3 standard deviations.  These are determined
by comparing the log of the likelihood to its peak value
\beq
\ln {\cal L} = \ln {\cal L}_{\rm peak} - s^2/2
\label{s}
\eeq
for $s$ standard deviations.
The contours in the figures represent 1, 2, and 3
standard deviations from the peak value as determined by (\ref{s}).
  If a large number of 
points contribute, the $s=2$  contour translates in 
the usual way to approximately 95\% confidence.

As an illustrative exercise it is useful first to consider a
sample of just one galaxy, the lowest metallicity galaxy I Zw 18
 where $Y$ has been measured independently in
 two distinct \hii regions, yielding
an average helium abundance 
\footnote{Recently, the smallness of the error bars associated with one
 of the two
\hii regions in I Zw 18 has been questioned due to underlying stellar
absorption (Skillman, Terlevich \& Terlevich 1997). Of course, too much weight
should never be given to any 
one single observation.}  $\langle Y_1,Y_2\rangle = .230 \pm .004$.
Because the measured values of $Y_p$ for the 
two \hii regions in  I Zw 18 are consistent with each other, 
the likelihood function (\ref{prob2}) is peaked at $w = 0$
and prefers a value $Y_p = 0.230 \pm 0.004$, equal to the
average of the two points (see figure 1a).

The full power of the method becomes clearer when we 
employ a larger statistical sample.
In   figures 1b-1e, we show the   effect of progressively
adding more
\he4 data to our sample.
  The data is ordered by metallicity
so that for example, the 11 point set includes the 11 extragalactic \hii 
regions with the lowest values of O/H.  Except
for sample selection, the oxygen abundance is   not explicitly
 utilized. We show the likelihood distributions up to what was described 
Olive, Skillman, and Steigman (1997) as set C corresponding to 32 data points.
In this case, the linear extrapolation with respect to O/H gave
$Y_p = 0.230 \pm 0.003$.  The cuttoffs for the various data sets considered 
here correspond to O/H $< $35, 45, 60, and 85 $\times 10^{-6}$ for
the 7, 11, 18, and 32 point sets respectively.\footnote{The ``full" 
data set of 62 
points is not used here.  Indeed, when one tries to use this data set,
with O/H as high as 145 $\times 10^{-6}$, the correlation
between $Y$ and O/H cannot be neglected. Even in the 32
point set considered here, the correlation is 
statistically significant.
Notice in table 1 that 
the likely value of $w$ increases from the 18 to the
32 point set;  larger samples contain more information
on enrichment but obscure the information on $Y_p$.}

As soon as one goes beyond a 
2-point data set, the dispersion in the data (normally associated
with enrichment correlated with metallicity) produces a likelihood
distribution which is peaked at a non-zero value for $w$---
yielding an estimate of the stellar helium enrichment of the sample. 
Consider for example the 11-point data set. The peak of the distribution
occurs at $w = 0.017$ and at that width, $Y_p = 0.228 \pm 0.003$. 
The individual helium abundances in this sample
range from 0.225 to 0.251, with a weighted mean  0.237 $\pm$ 0.002.
The value for $Y_p$ determined by the likelihood distribution
is significantly below the mean value of the data.
This bias  
is familiar from studies of the Malmquist effect: the 
best estimate of the distribution from which   points are drawn
is in general quite different from the distribution of  the 
values for the points themselves. In the present context it is the 
asymmetry of the prior which leads to the apparent
 bias.

It is interesting to note that
at the 2$\sigma$ level, the data are consistent with a single primordial value
with no subsequent enrichment $(w=0)$.
This corresponds to a conservative 2$\sigma$ upper limit, $Y_p\le
0.240.$
As the number of data points in the sample  increases,
the peak of the likelihood distribution
shifts to higher values of $Y_p$. The 2$\sigma$ upper limit 
however remains at
$w = 0$, and its value is relatively insensitive to the data set.
These results are summarized in table 1.

The most important arbitrary step in the above analysis is in
the choice of prior, and to calibrate its effect we
consider a range of physically plausible possibilities.
In fact   the 
top-hat function is already quite conservative 
towards allowing the largest possible  
upper limits on $Y_p$--- that is, since very low $Y$ 
systems in the present Universe, even in the current samples
of low metal galaxies, are quite rare, it appears that
stellar helium enrichment is seldom very small. 
We could however try to bias the result  by choosing a
different shape biased {\it a priori} towards  smaller (or larger)
nonprimordial enrichments.  
Consider for example a sawtooth function
 prior, $P(Y_{iT} | Y_p) = (Y_p + w - Y_{iT})/w $
for $Y_{iT}$ between $Y_p$ and $Y_p + w$ and 0 otherwise.
This stipulates   {\it a priori}  
that  it  is most likely that there
has been very little helium enrichment, and no enrichment greater than
$w$, thus $P$ rises abruptly to some value at $Y_{iT}= Y_p$,
linearly decreasing for  $ Y_p\le Y_{iT}\le Y_p + w$,
and zero for $Y_{iT}\ge Y_p + w$. We will refer to this prior as 
the negative bias (named for the slope with respect to $Y_{iT}$).
Alternatively, we might
 choose   $P(Y_{iT} | Y_p) = (Y_{iT} - Y_p)/w $
for $Y_{iT}$ between $Y_p$ and $Y_p + w$ and 0 otherwise.
This assigns a greater probability for larger  
helium enrichment but still no enrichment greater than
$w$; thus $P$ rises linearly from $Y_p$ to some value at $Y_{iT}= Y_p + w$.
This is indeed the case for
typical galaxy samples, where low enrichment
systems are hard to find, so this is perhaps the more realistic prior.
We   refer to this prior as 
the positive bias. Taken together, these priors span the range of
reasonable possibilities.

Likelihood contours  for these two additional choices for priors are
 shown in   figures 2 and 3 for the same subsamples discussed above. 
 For all three priors, and for all the samples considered, the
goodness of fit of the model--- the effective reduced 
$\chi^2$--- is less than unity, reflecting that the model
is an adequate description of the data (and that the random observational 
errors have typically been overestimated.)  The differences in the 
parameter estimates 
from the top-hat case are small, reflecting 
the fact that the most critical assumption affecting the results
is the one we are most confident in, namely  the
assignment of $P(Y_{iT} | Y_p)=0 $ for  $   Y_{iT}\le Y_p$.
As is evident from the figures, the likelihood contours are
shifted to the right for the negative bias prior, yielding a higher value for
$Y_p$ and to the left for the positive bias,  favoring a lower $Y_p$.
However, the $2\sigma$ upper limit is found to be nearly independent
of  our choice of prior.  These results are summarized in tables 2 and 3.

\section{Conclusions}

The significant detection of nonzero $w$ reveals 
 that even in these samples of metal-poor galaxies,
there is   evidence for a spread in $Y$ produced by stars.
The best  fit to the data is for values of the primordial 
abundance in the range $0.221\le Y_p\le 0.236$, depending on
the specific model and subsample.  A 2$\sigma$
upper limit, $Y_p\le 0.243$, holds for all cases;
this bound is  virtually independent of the stellar helium
enrichment model. There are enough data points
entering so that even if individual errors are nongaussian,
this limit corresponds approximately to a 95\% confidence level.

The statistical evidence from this sample indicates
 that the internally estimated
errors are if anything overly conservative, so 
the quoted limits are also generously estimated.
The only reasonable way
to reconcile the data with a higher value of $Y_p$ would be 
some systematic error in common among all the points--- that is, 
an in-common mistake biasing the results of all the   measurements. 
Of course, errors of this kind cannot be corrected by any purely  statistical 
technique.

Even with quite simple and  conservative assumptions, our analysis yields a
final limit which is sufficiently precise to overconstrain the 
Big Bang picture when combined with other data.
Our limit of $Y_p\le 0.243$ corresponds to a limit on the baryon/photon ratio
$\eta\le 3.5\times10^{-10}$ and a predicted deuterium abundance
$D/H\ge 6.2\times 10^{-5}$, in conflict with some recent claims
(e.g. Tytler \etal 1996) but in accord with others
(e.g. Songaila \etal 1996). It is also consistent with BBN predictions
of the Li abundance (Fields \& Olive 1996, Fields \etal 1996).
For the observed 
microwave background temperature, the baryon density is predicted
to be
$\Omega_bh^2\le 0.013$,
which begins to conflict with some recent estimates based on models
of quasar Lyman-$\alpha$
 absorption (e.g., Rauch \etal 1997, Weinberg \etal 1997), but 
 accords with other estimates of baryon density (e.g., Hogan 1997ab).
An overall concordance of the Big Bang picture is certainly possible
but will depend on which of these datasets ``gives way''.  It is
clear that the reliability of $Y_p$ estimates would improve significantly
with comprehensive studies of  even one more region similar to I Zw 18.

\acknowledgements
We would like to thank E. Skillman and G. Steigman for helpful conversations.
We are grateful to the Institute for Nuclear Theory, Seattle,
funded by DOE, 
for sponsoring the 1996 workshop on Nucleosynthesis in the Big Bang,
Stars, and Supernovae, where this work was started. This work
was also supported at the University of Washington by NASA and NSF, and at
the University of Minnesota in part by  DOE grant DE-FG02-94ER-40823.

\newpage
\beginapjbib




\bibitem Fields, B.D. \& Olive, K.A. 1996, Phys Lett B368, 103

\bibitem Fields, B.D., Kainulainen, K., Olive, K.A., \& Thomas, D. 1996
New Astronomy, 1, 77



\bibitem Hogan, C. J., 1997a, in ``Critical Dialogues in Cosmology''
ed. N. Turok (Princeton: Princeton University Press; astro-ph/9609138)

\bibitem Hogan, C. J. 1997b, in Proceedings of the 18th Texas
Conference on Relativistic Astrophysics,
ed. A. Olinto, J. Frieman, and D. Schramm,
(World Scientfic, in press) (astro-ph/9702044) 

\bibitem Hogan, C. J., Anderson, S. F. and Rugers, M. H. 1997,
AJ, in press (astro-ph/9609136)

\bibitem Izotov, Y.I., Thuan, T.X., \& Lipovetsky, V.A. 1994
ApJ 435, 647.  

\bibitem Izotov, Y.I., Thuan, T.X., \& Lipovetsky, V.A. 1997, ApJS, 108, 1


\bibitem Kunth, D., Lequeux, J., Sargent, W.L.W., \& Viallefond, F. 
1994, A\&A, 282, 709



\bibitem Olive, K.A. \& Scully, S.T. 1996, IJMPA, 11, 409

\bibitem Olive, K.A., Skillman, E., \& Steigman, G. 1997, ApJ, 483, in press

\bibitem Olive, K.A., Steigman, G. \& Walker, T.P. 1991, ApJ, 380, L1

\bibitem Olive, K.A., \& Steigman, G. 1995, ApJS, 97, 49

\bibitem Pagel, B E.J., Simonson, E.A., Terlevich, R.J.
\& Edmunds, M. 1992, MNRAS, 255, 325 (PSTE)




\bibitem Peimbert, M. \& Torres-Peimbert, S. 1974 Ap J, 193, 327

\bibitem Press, W. H. 1996. astro-ph/9604126

\bibitem   Rauch, M., \etal 1997, submitted to ApJ (astro-ph/9612245)


\bibitem  Sarkar, S., 1996, Rep. Prog. Phys, 59, 1493




\bibitem Skillman, E., \& Kennicutt 1993, ApJ, 411, 655 

\bibitem Skillman, E., Terlevich, R.J., Kennicutt, R.C., Garnett, D.R.,
\& Terlevich, E. 1994, ApJ, 431, 172 

\bibitem Skillman, E., Terlevich, R.J., Telrevich, E. 1997, in preparation

\bibitem Skillman, E., \etal 1997, (in preparation)  

\bibitem Smith, M. S., Kawano, L. H., and Malaney, R. A. 1993, ApJ
85, 219

\bibitem Songaila, A., Wampler, E.J., \& Cowie L.L. 1997, Nature, 385,
127



\bibitem Tytler, D., Burles, S., \& Kirkman, D. 1996,
submitted to the Astrophysical Journal (astro-ph/9612121)


 \bibitem Walker, T. P., Steigman, G., Schramm, D. N., Olive, K. A.,
 \& Kang, H. 1991, ApJ, 376, 51

\bibitem Weinberg, D.,   Miralda-Escude, J.,  Hernquist, L.,
and   Katz, N. 1997, submitted to ApJ (astro-ph/9701012)
\endapjbib
\clearpage
\centerline{\bf Tables}
\bigskip
\centerline{Table 1: Likeliest values and limits for the top-hat prior}
\begin{center}
\begin{tabular}{|cccc|}  \hline \hline                   
\# Regions  & $w$ &  $Y_{p}$  & $2\sigma$ upper bound \\ \hline
2 & 0 & 0.230 & 0.237 \\
7 & 0.021 & 0.226 & 0.238 \\
11 & 0.017 &0.228 & 0.240 \\
18  & 0.011 & 0.232 & 0.241 \\ 
32 & 0.013 & 0.234 & 0.242 \\
\hline
\end{tabular}
\end{center}
 
\bigskip

\centerline{Table 2:  Likeliest values and limits for the negative  bias}
\begin{center}
\begin{tabular}{|cccc|}  \hline \hline                   
\# Regions  & $w$ &  $Y_{P}$  & $2\sigma$   upper bound \\ \hline
2 & 0 & 0.230 & 0.237 \\
7 & 0.025 & 0.228 & 0.238 \\
11 & 0.021 &0.230 & 0.240 \\
18  & 0.013 & 0.234 & 0.241 \\ 
32 & 0.015 & 0.236 & 0.243 \\
\hline
\end{tabular}
\end{center}
\bigskip
\centerline{Table 3:  Likeliest values and limits for the positive bias}
\begin{center}
\begin{tabular}{|cccc|}  \hline \hline                   
\# Regions  & $w$ &  $Y_{P}$  &  $2\sigma$ upper bound \\ \hline
2 & 0 & 0.230 & 0.237 \\
7 & 0.024 & 0.221 & 0.239 \\
11 & 0.020 &0.224 & 0.239 \\
18  & 0.014 & 0.229 & 0.241 \\ 
32 & 0.016 & 0.230 & 0.242 \\
\hline
\end{tabular}
\end{center}
\bigskip

\clearpage

\beginfig

\figitem{\bf Figure 1:} {Likelihood   function  
showing $1\sigma$, $2\sigma$ and $3\sigma$ contours in the 
$(Y_p,w)$ plane,
for a top-hat prior of width $w$. The +'s indicate
the peaks of the likelihood functions. Results for different subsamples
are shown in panels 1a-1e, starting with the 2 points of I Zw 18
and ending with the 32 lowest metal points.}

\figitem{\bf Figure 2:} { Same as Figure 1, for the negative bias prior.}

\figitem{\bf Figure 3:} { Same as Figure 1, for the positive bias prior. }
\endfig

\begin{figure}[htb]
\hspace{-1truecm}
\epsfysize=8.0truein
\epsfbox{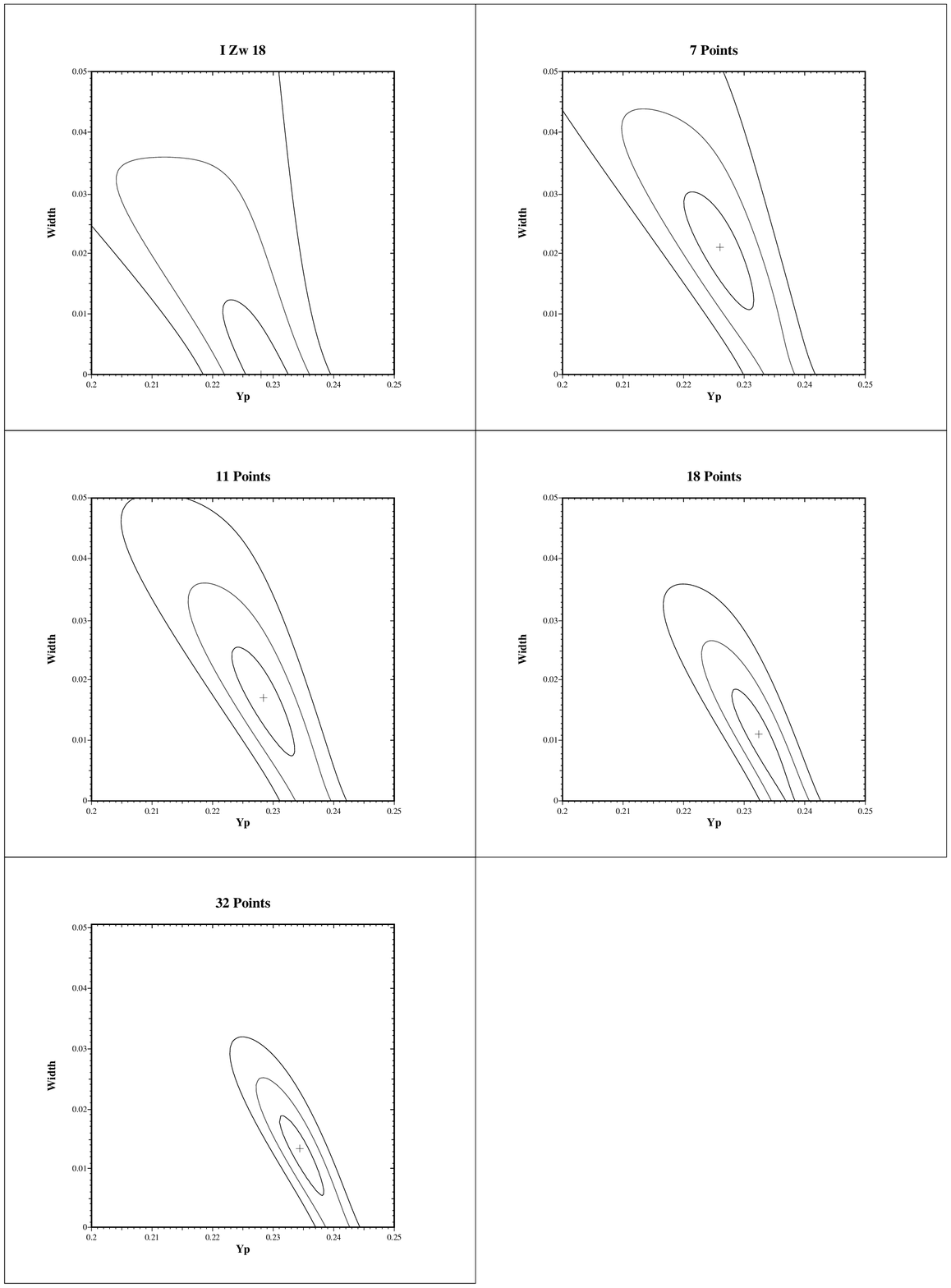}
\end{figure}  

\newpage

\begin{figure}[htb]
\hspace{-1truecm}
\epsfysize=8.0truein
\epsfbox{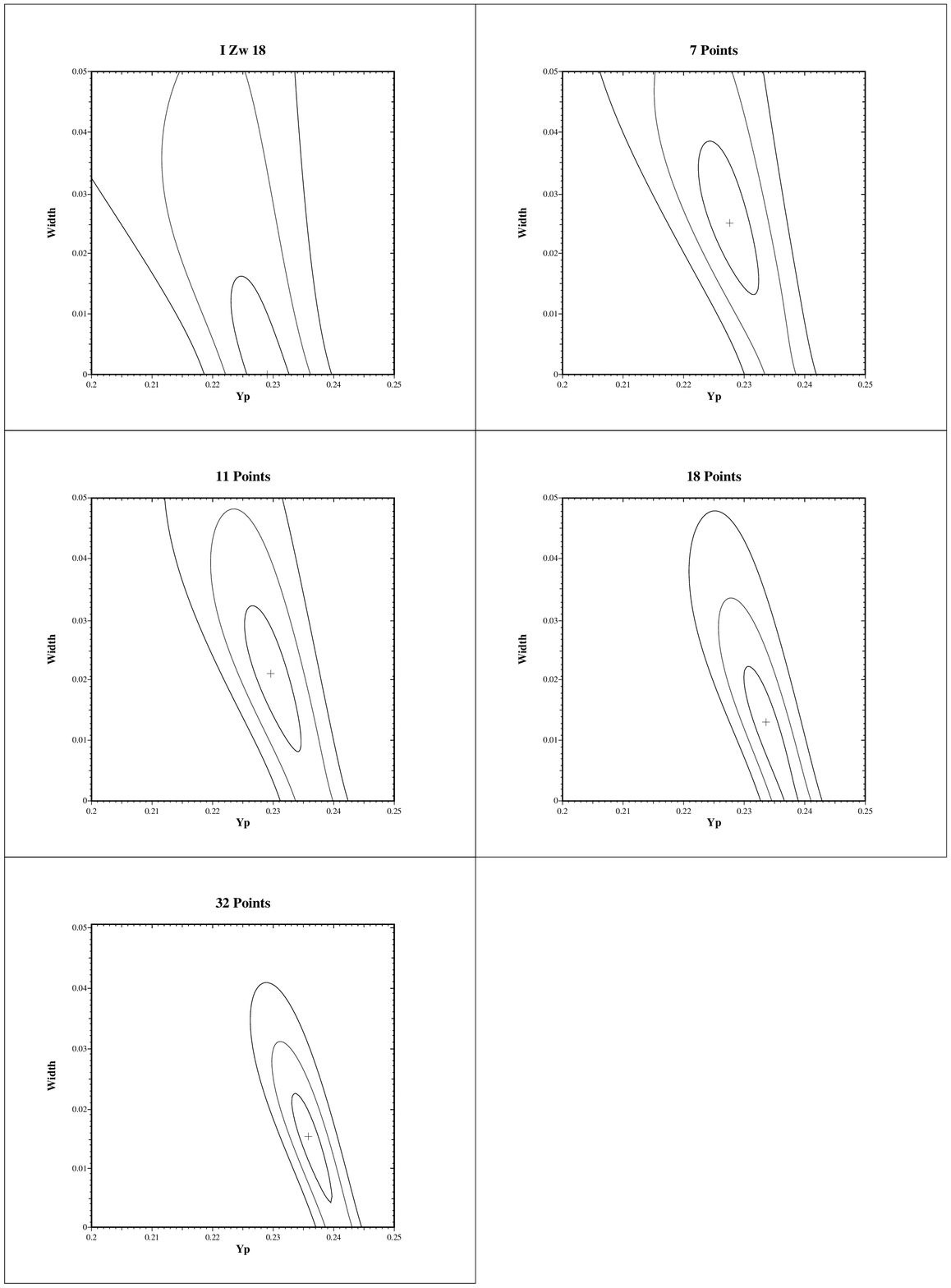}
\end{figure}

\newpage

\begin{figure}[htb]
\hspace{-1truecm}
\epsfysize=8.0truein
\epsfbox{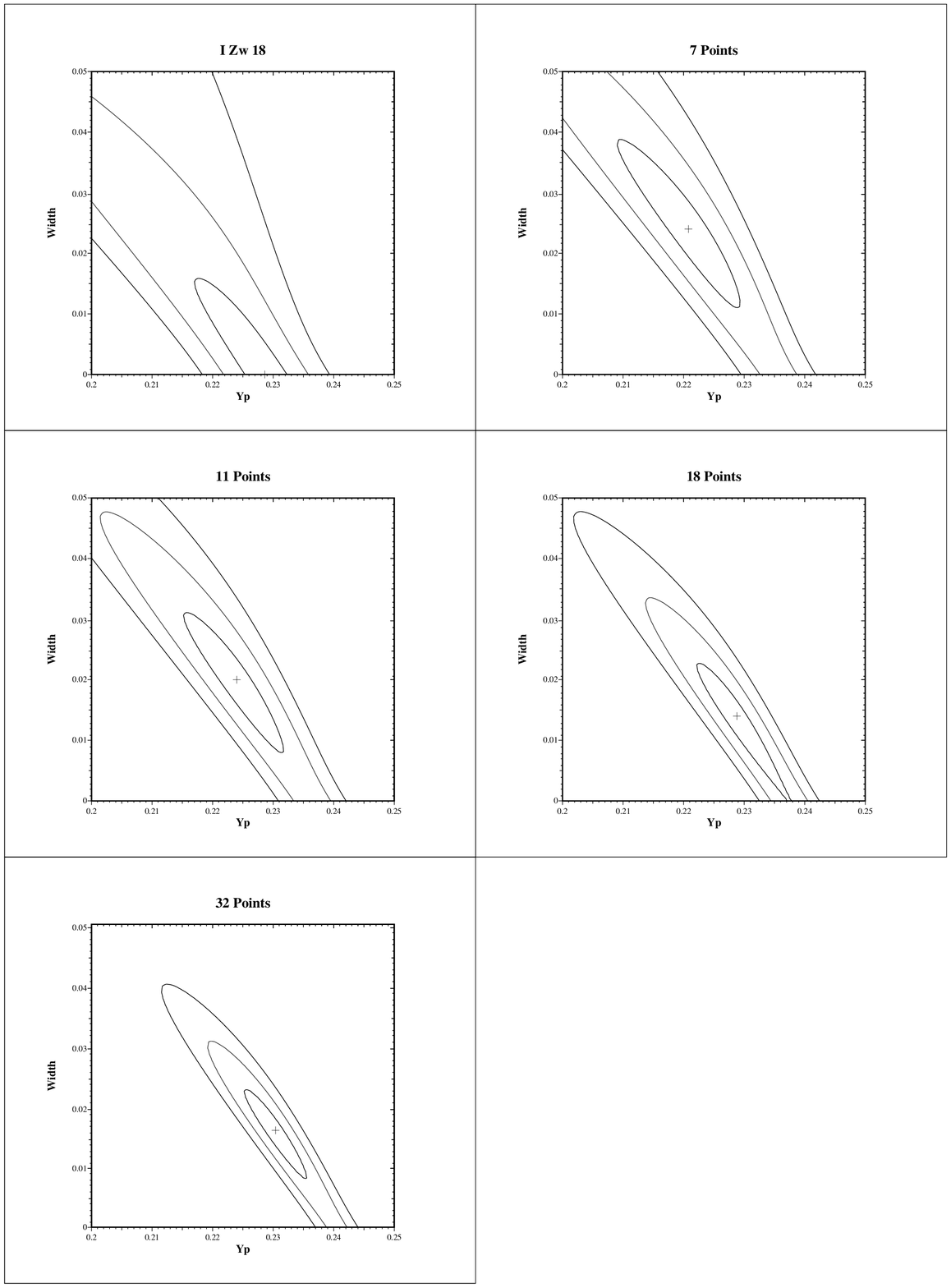}
\end{figure}

\end{document}